\documentclass[prb,twocolumn]{revtex4}

     \usepackage{bm}
     \usepackage{graphicx}
     \usepackage{amssymb}
     \usepackage{amsmath}
     \usepackage{eufrak}
     \usepackage{color}
     \usepackage[utf8]{inputenc} 
     \usepackage{hyperref}
     \usepackage{pifont}
     \usepackage{ulem}
     \usepackage{verbatim}

\allowdisplaybreaks
			
\newcommand{\nix}[1]{}
\begin{document}

\title{Magneto-resistance oscillations
induced by high-intensity terahertz radiation}

\author{
T.~Herrmann,$^1$ Z.~D.~Kvon,$^{2,3}$ I.~A.~Dmitriev,$^{1,4}$ D.~A.~Kozlov,$^{2,3}$ B.~Jentzsch,$^1$  M.~Schneider,$^1$ L.~Schell,$^1$
V.~V.~Bel'kov,$^4$ A.~Bayer,$^1$ D.~Schuh,$^1$ D.~Bougeard,$^1$  T.~Kuczmik,$^1$ M.~Oltscher,$^1$ D.~Weiss,$^1$
and S.~D.~Ganichev$^{1}$
}

\affiliation{$^1$ Terahertz Center, University of Regensburg, 93040 Regensburg, Germany}
\affiliation{$^2$ Rzhanov Institute of Semiconductor Physics, 630090 Novosibirsk, Russia }
\affiliation{$^3$ Novosibirsk State University, 630090 Novosibirsk, Russia }
\affiliation{$^4$ Ioffe Institute, 194021 St.\,Petersburg, Russia}

\begin{abstract}
We report on observation of pronounced terahertz radiation-induced magneto-resistivity oscillations in AlGaAs/GaAs two-dimensional electron systems, the THz analog of the microwave induced resistivity oscillations (MIRO). Applying high power radiation of a pulsed molecular laser we demonstrate that MIRO, so far observed at low power only, are not destroyed even at very high intensities. Experiments with radiation intensity ranging over five orders of magnitude from 0.1 W/cm$^2$ to 10$^4$ W/cm$^2$ reveal high-power saturation of the MIRO amplitude, which is well described by an empirical fit function $I/(1+I/I_s)^\beta$ with $\beta~\sim 1$. The saturation intensity $I_s$ is of the order of tens of W/cm$^2$ and increases by six times 
by increasing the radiation frequency from 0.6 to 1.1~THz.
The results are discussed in terms of microscopic mechanisms of MIRO and compared to nonlinear effects observed earlier at significantly lower excitation frequencies.
\end{abstract}

\maketitle{}

Magnetotransport experiments in low-dimensional systems containing high mobility two-dimensional electron gases (2DEG) reveal many fundamental phenomena of quite different physical nature. The most prominent and well-known examples in linear dc transport are integer and fractional quantum Hall effects~\cite{klitzing:1980,tsui:1982} in stronger magnetic field and  Shubnikov - de Haas  (SdH)~\cite{ando:1982,shoenberg:1984} and Weiss~\cite{weiss:1989}
oscillations at moderate fields. While linear transport phenomena in low-dimensional semiconductor systems have been systematically studied for several decades, in the last years terahertz/microwave-induced nonequilibrium transport in 2DEG is attracting an ever growing attention. In part, this is caused by the steadily expanding frequency range and rapid developments of terahertz science and technology~\cite{Ganichevbook,Dexheimerbook,Leebook,Zhangbook,Mittlemanbook,Bruendermannbook}. Following the discovery of the microwave-induced resistance oscillations (MIRO) \cite{zudov:2001,ye:2001} and associated zero-resistance states \cite{mani:2002,zudov:2003,yang:2003,dorozhkin:2003,smet:2005}, the focus of recent research has largely shifted to non-equilibrium magnetotransport phenomena \cite{dmitriev:2012}. Similar to the SdH and Weiss oscillations, MIRO are $1/B$-periodic oscillations and reflect the commensurability between the photon energy $2\pi \hbar f$ and the cyclotron energy $\hbar \omega_c$. Here, $\hbar$ is the reduced Planck constant, $f$ the microwave frequency and $\omega_c$ the cyclotron frequency. Very recently, it was demonstrated that MIRO can be efficiently excited at substantially higher frequencies of several terahertz~\cite{herrmann:2016}. This work gave an experimental answer to two currently most intriguing questions regarding radiation helicity \cite{smet:2005} and the role of the contacts/edges \cite{mikhailov:2011,chepelianskii:2009}. So far all studies of MIRO were performed at low radiation power smaller or of the order of a milliwatt~\cite{dmitriev:2012}.

Here we demonstrate that MIRO are very robust and do not vanish even at very high power up to tens of kW/cm$^2$. We have studied MIRO for frequencies between 0.6 and 1.1~THz and intensities varying by five orders of magnitude. We observe that high radiation intensity affects exclusively the amplitude of MIRO while both shape and phase of the oscillations are preserved. For all frequencies and all oscillation orders the dependence of the MIRO amplitude on intensity $I$ follows an empirical fit function $I/(1+I/I_s)^\beta$ with $\beta~\sim 1$. The saturation intensity $I_s$ strongly depends on excitation frequency and is largely independent of the oscillation order. We find that a bolometric response of the non-oscillating magneto-resistivity background
shows different intensity dependence and enters the nonlinear regime at substantially lower intensities. The results are discussed in terms of microscopic mechanisms of MIRO and compared to high-power effects observed earlier at much lower excitation frequencies.

\section{Samples and experimental technique}

The experiments are carried out on doped (001)-oriented molecular-beam
epitaxy (MBE) AlGaAs/GaAs quantum well (QW) structures.
The samples are prepared in Corbino disk geometry, which measure
directly the longitudinal conductivity $\sigma_{xx}$. The transport parameters of the samples as well as the inner, $r_i$, and outer, $r_o$,
radii of the gold-germanium contacts are summarized in Table\,\ref{sample}.
Note that samples \#A,  \#B, and \#D have been studied in our work on
MIRO excited by low power terahertz radiation \cite{herrmann:2016}
and are labeled in the same way as there.
We can thus directly compare the MIRO behavior in linear and nonlinear regimes.
In order to explore the effect of the sample geometry, we prepared samples \#G$_S$,  \#G$_M$,  and
\#G$_L$ from the same wafer such that they differ in the radius of the inner contact only.
The corresponding experimental results are discussed in Appendix~\ref{appendixgeometry}.
Most of the data shown here are taken after the cooled samples
have been exposed to room light. Under illumination
MIRO become much more pronounced, as illustrated by comparison of the data
obtained on illuminated and non-illuminated structures in Appendix~\ref{appendixillum}.

\begin{table}
\centering
\begin{tabular}{|c|c|c|c|c|c|}
	\hline
	  Sample          &$r_i$ &$r_o$ &$\mu$              &$n_e$        & QW thickness        \\
                    &  [mm]  &   [mm] &10$^3$\,[cm$^2/$Vs]&10$^{11}$\,[cm$^{-2}$] & [nm]\\
\hline
		\#A
							& 0.25 & 4.25  & 820               & 12.0        & 10        \\
	  \#B
								& 0.3  & 1.0  & 1800              & 9.3         &12.5      \\
		 \#D$_S$
							& 0.25 & 1.5  & 980               & 24.0        &10    \\
		 \#D$_L$
							& 0.25 & 4.25  & 980               & 24.0        &10    \\
	
	  \#G$_S$
								& 0.5  & 4.25  & 680               & 13.5        &10      \\
		\#G$_M$
							& 1.0 & 4.25  & 680               & 13.5        &10     \\

		 \#G$_L$
							& 1.5 & 4.25  & 680               & 13.5         &10    \\
		\hline
\end{tabular}
\caption{Sample parameters and transport data obtained at $T=2$\,K  including the electron density $n_e$ and mobility $\mu$.
}
\label{sample}
\end{table}

\begin{figure}[h]
\includegraphics[width=1.0\linewidth]{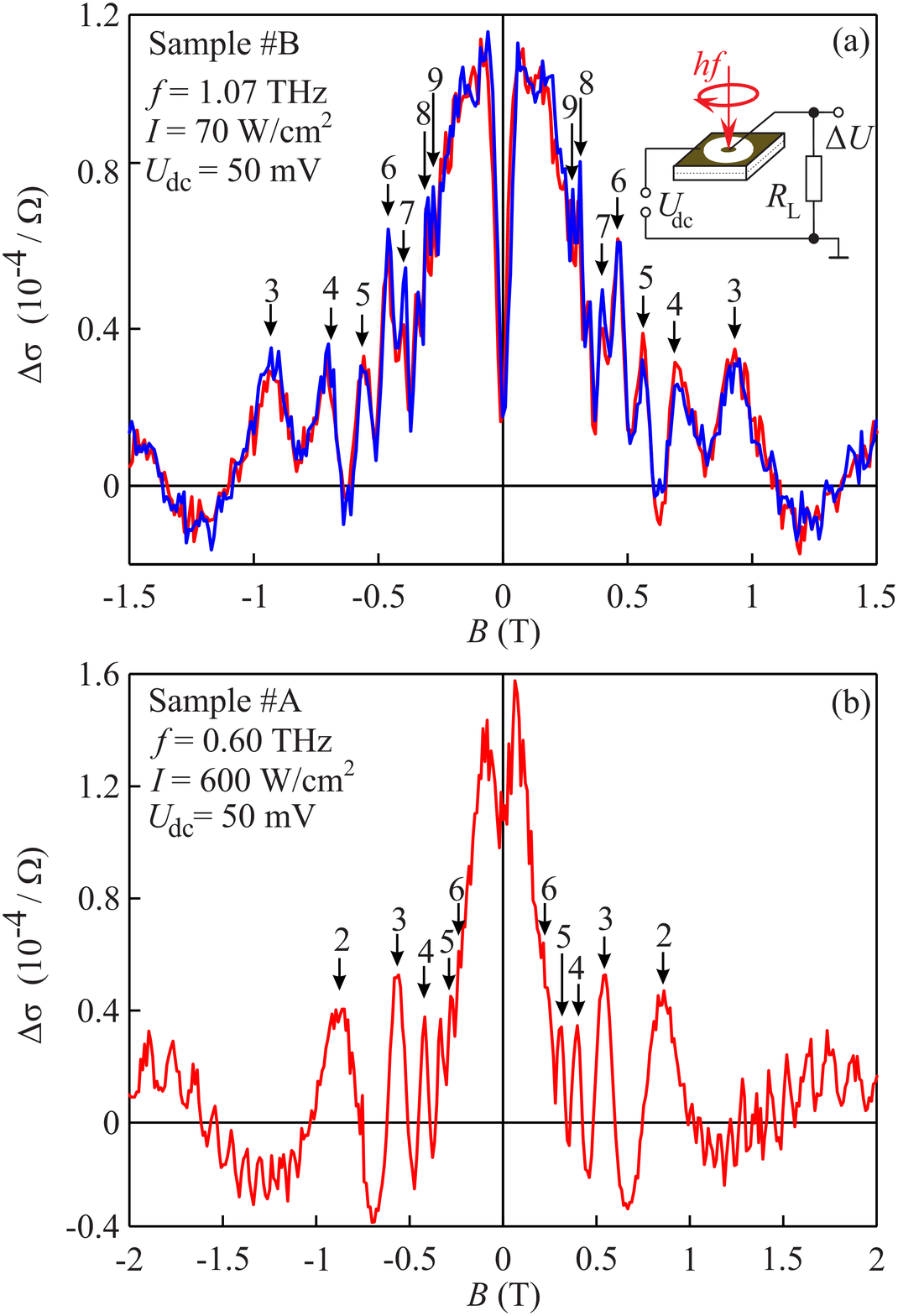}
\caption{Magnetic field dependence of the photoconductivity
$\Delta\sigma$  induced by $f = 1.07$ and $0.6$~THz
pulsed radiation in samples \#B and \#A,  see legend.
Oscillation maxima are marked by arrows and numbers denote
the closest integer $\epsilon = \omega / \omega_c$ (oscillation order).
Red and blue lines in panel (a) correspond to the data obtained for
right- and left-handed circularly polarized radiation, respectively,
while inset sketches the experimental set-up.
}
\label{Fig1}
\end{figure}

To excite MIRO we apply a high-power line-tunable  molecular THz laser~\cite{Ganichev93,Kvon08,Schneider04,Lechner2009}
optically pumped by a pulsed transversely excited atmosphere (TEA) CO$_2$ laser~\cite{SGEopt2003,Chongyun}.
The laser operated at
frequencies $f = 0.60$, $0.78$,
and $1.07$\,THz, which were obtained with
NH$_3$,
D$_2$O, and
CH$_3$F gases, respectively.
The corresponding photon energies ($\hbar \omega = 2.5$, 3.2, and 4.4\,meV, respectively)
are smaller than the
energy distance to the second size-quantized subband.
Consequently, the radiation induces indirect optical transitions (Drude-like free-carrier absorption)
in the lowest subband.
The laser generated single pulses with a duration of about 100\,ns and
a repetition rate of 1\,Hz, which were focused onto a spot size of about $2.5$\,mm
diameter, depending on the radiation frequency.
The laser peak power
was controlled by the THz photon drag detector~\cite{drag}.
The radiation intensity was varied from 10$^{-1}$ to 10$^4$ W/cm$^2$
using a set of calibrated teflon, polyethylene and pertinax attenuators~\cite{Ganichevbook}.
The beam had an almost Gaussian profile which was measured by a pyroelectric camera\,\cite{Ganichev1999,Ziemann2000}.
Right- and left-handed circularly polarized radiation were achieved by
transmitting the linearly polarized 
laser
beam through $\lambda$/4 crystal quartz plates\,\cite{BelkovSSTlateral,Kohda2012,ratchet2011}.
The photoconductive response for normally incident radiation was measured as a function of
magnetic field $\bm B$ up to 7\,T applied perpendicularly to the QW plane.
A sketch of the setup is shown  in the inset in Fig.\,\ref{Fig1}~(a). 
To obtain the photoconductivity $\Delta\sigma$, we measured the THz radiation-induced voltage  drop
$\Delta U$ across the load resistor $R_L = 50$\,Ohm at a fixed bias voltage $U_{\rm dc}=50$\,mV.
The signals were fed into amplifiers (voltage amplification by
a factor of 100  and a bandwidth of 300~MHz) and were recorded
by a digital  broad-band (1~GHz) oscilloscope.
All experiments were performed at the liquid helium temperature $T=4.2$~K.

\section{Experimental results}

Typical experimental traces of photoconductivity for samples \#A and \#B are shown in Fig.~\ref{Fig1}.
They feature strong $1/B$-magnetooscillations (MIRO) extending to $B$ below 0.3 T,
on top of a nonmonotonous background. The periodicity of MIRO reflects the ratio $\epsilon=\omega/\omega_c$
of the radiation angular frequency $\omega=2\pi f$ and the cyclotron frequency $\omega_c=e B/m_e$.
Arrows in Fig.\,\ref{Fig1} mark the maxima of the oscillations and the
numbers denote the closest integer $\epsilon$, called the oscillation order in what follows.
In sample \#A, see Fig.~\ref{Fig1}~(b), one also observes the radiation-induced changes in the Shubnikov-de Haas oscillations at higher $B>1$~T.
The visibility of SdH in the photoconductivity for all samples is consistent with their
presence in corresponding transport measurements without radiation (dark conductivity, not shown).

\begin{figure}
\includegraphics[width=0.9\linewidth]{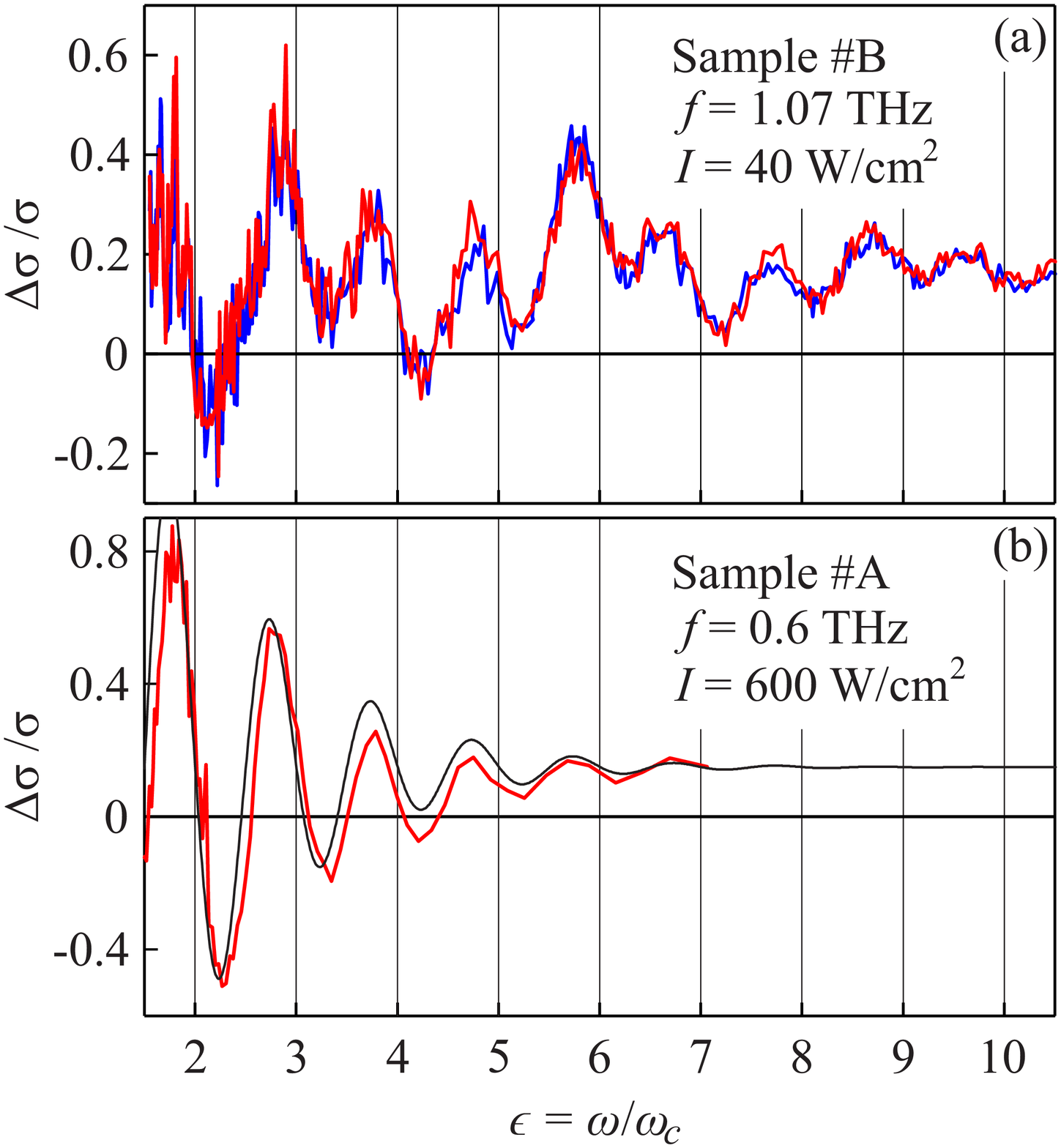}
\caption{Photoconductivity $\Delta \sigma$ normalized  to dark conductivity
$\sigma$ as a function of $\epsilon = \omega / \omega_c$. Panels (a) and (b)
 present data for samples \#B and \#A illuminated by radiation of frequency $f = 1.07$ and $0.6$ THz,
respectively. In addition, the thin black line in panel (b) presents the fit function
 $\Delta\sigma/\sigma=0.15 - 3.5 \exp(-\epsilon/f\tau_q)\epsilon\sin(2\pi\epsilon)$.
 Here the damping parameter $\tau_q=1.5$~ps is taken from similar fitting of the low-intensity data,
 obtained for $I\simeq 0.11$~W/cm$^2$ on the same sample in Ref.~\onlinecite{herrmann:2016}. The values of the effective mass $m_e$
 for both samples, needed to fix the scaling of the $\epsilon=2\pi m_e f/e B$, are also taken from Ref.~\onlinecite{herrmann:2016}.
 }
\label{Fig2}
\end{figure}

Figure~\ref{Fig2} shows the ratio $\Delta\sigma/\sigma$, i.e. the photoconductivity $\Delta\sigma$ normalized to the independently measured dark conductivity $\sigma$,
as a function of $\epsilon$. 
In the region where MIRO are observed, both $\sigma$ and the non-oscillatory part of $\Delta\sigma$ scale roughly as $B^{-2}$
as expected for Corbino samples in the regime of classically strong magnetic field, $\mu B\gg 1$. This is clearly seen in Fig.~\ref{Fig2},
where the non-oscillatory background of the ratio $\Delta\sigma/\sigma$ is indeed nearly $B$-independent. Using this property, we can easily separate 
 the oscillatory, $\Delta\sigma_\text{osc}$, and background, $\Delta\sigma_\text{bg}$, contributions to the photoconductivity $\Delta\sigma=\Delta\sigma_\text{bg}+\Delta\sigma_\text{osc}$.
The value of electron mass $m_e$, required for a proper scaling of the $\epsilon$-axis in Fig.~\ref{Fig2},
was taken from our previous work on the same sample, see Ref.~\onlinecite{herrmann:2016}. There, it was determined both from the cyclotron resonance position in transmission measurements and from the periodicity of MIRO at $f=0.69$~THz.  Within the experimental accuracy, these two methods provided an identical value of $m_e=0.074~m_0$ (which happens to be the same for samples \#A and \#B).

The data in Fig.~2 demonstrate perfect agreement with basic MIRO phenomenology established before \cite{dmitriev:2012}. In particular, the minima and maxima are found at $\epsilon = N \pm 1/4$, symmetrically offset from the nodes at integer $\epsilon=N=1,\,2,\,3\ldots$. The photoresponse at $\epsilon=N$ can thus be used to separate a nearly $B$-independent background photoconductivity $\Delta\sigma_\text{bg}$ from the oscillating signal $\Delta\sigma_\text{osc}$.
The data for right- and left-handed circularly  polarized radiation presented in
Fig.\,\ref{Fig1}\,(a) and~\ref{Fig2}\,(a), i.e. obtained under conditions corresponding to
the cyclotron resonance active (CRA) and inactive\,(CRI) configurations,
show almost indistinguishable traces of MIRO.

\begin{figure}
\includegraphics[width=0.9\linewidth]{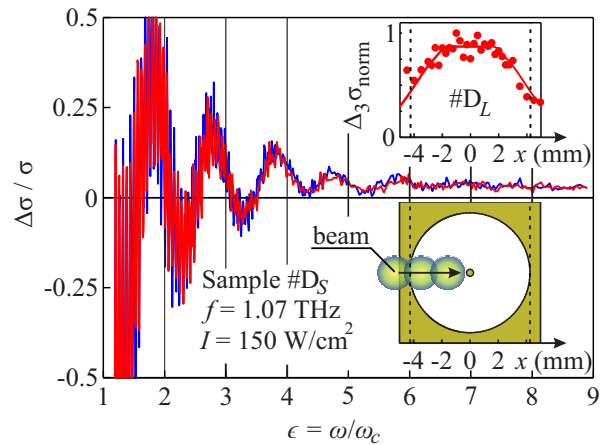}
\caption{Normalized photoconductivity $\Delta \sigma / \sigma$ as a
function of $\epsilon = \omega / \omega_c$ for sample \#D$_S$
illuminated by $f = 1.07$~THz laser pulses. 
Red and blue curves correspond to the right- and left-handed
circularly polarized radiation.
The upper inset shows the MIRO amplitude $\Delta_3\sigma_\text{norm} = \Delta_3\sigma (x)/\Delta_3\sigma_\text{max}$  for oscillation order $\epsilon =3$ as a function 
of the laser spot position $x$. 
Here $\Delta_3\sigma_\text{max}$ is  the  maximal value of $\Delta_3\sigma (x)$. These data were obtained for sample \#D$_L$ and $I=500$~W/cm$^2$.  
The measurement setup is shown in the lower inset: The beam spot with a radius $d/2 \approx 1.25$~mm being smaller
than $r_o = 4.25$~mm  (but larger than $r_i = 0.25$~mm) is scanned across the Corbino disk.
The line in the upper inset is a guide for the eye. 
}
\label{Fig3}
\end{figure}

To prove that photoconductive signal does not come from irradiating the edges we
scanned the beam across the sample \#D, as shown in the lower inset to Fig.\,\ref{Fig3}.
The upper inset in this figure shows that the amplitude of MIRO (defined below) is maximal when the laser beam is close to the sample center and
decreases as the beam approaches Corbino disk edges. This dependence demonstrates that the oscillation amplitude
just follows the decrease in absorbed radiation power.

\begin{figure}
\includegraphics[width=0.9\linewidth]{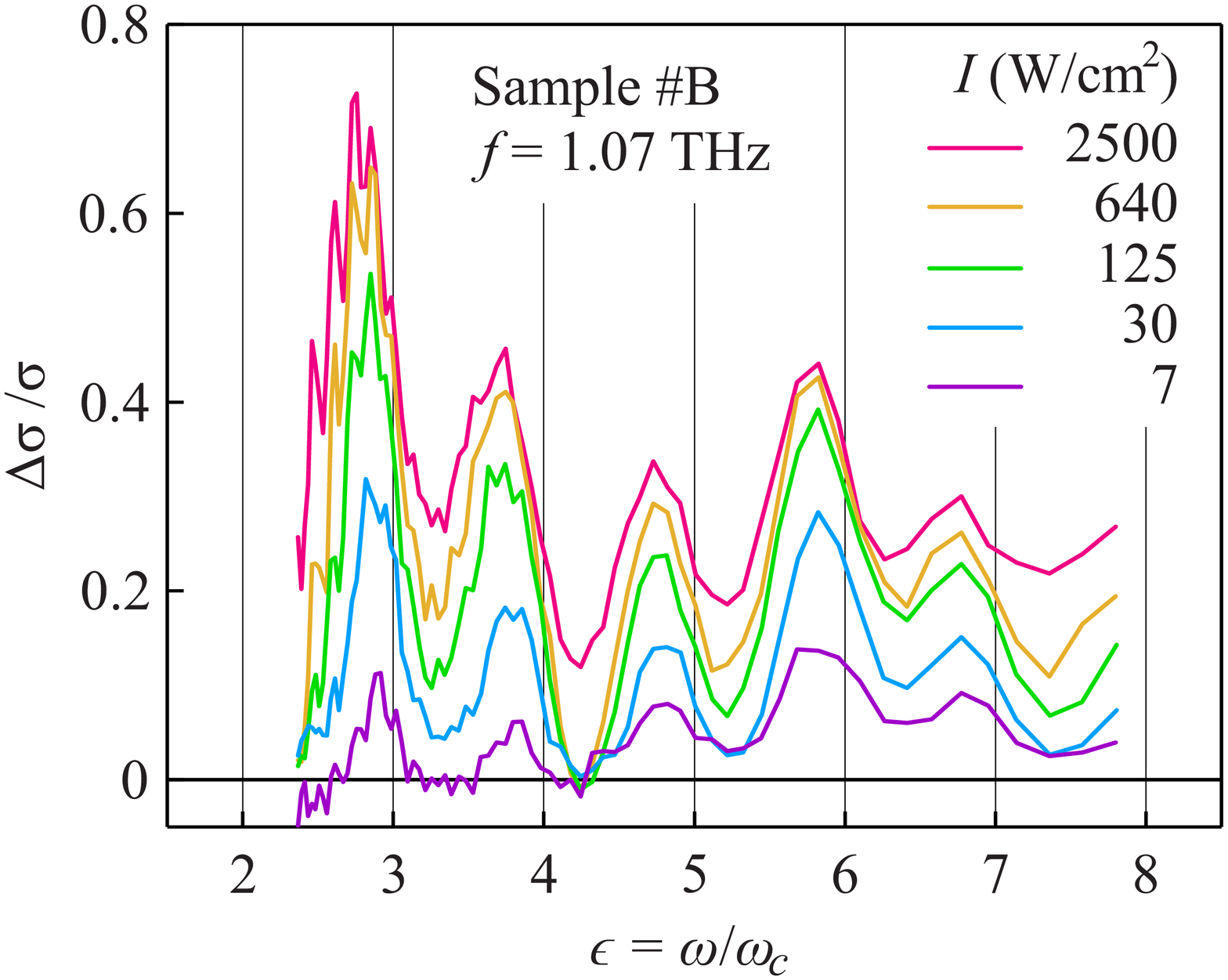}
\caption{Normalized photoconductivity $\Delta \sigma / \sigma$
as a function of $\epsilon = \omega / \omega_c$ for sample \#B
excited by $f = 1.07$ THz radiation with of different intensity as marked.
}
\label{Fig4}
\end{figure}

The overall behavior of the $1/B$--oscillations in Figs.~\ref{Fig1}--\ref{Fig3}, including the exponential damping at low $B$ and the $1/4$-shift of the minima and maxima from the nodes, is typical for the linear regime of MIRO, $\Delta\sigma\propto I$.
At the same time, it is quite remarkable that in our experiments MIRO preserve their shape, phase, and even low-$B$ damping
in a wide range of intensities from $0.1$ up to $10^4$~W/cm$^2$, such that only the overall
amplitude depends on the intensity of radiation. Such behavior is detected for all samples, frequencies, and
radiation intensities, and is illustrated in Fig.~\ref{Fig4} showing several traces of MIRO for sample \#B  at different
intensities of the $f=1.07$~THz radiation. Despite this particular sample demonstrates somewhat irregular shape of MIRO,
this shape is well reproduced for all intensities, such that the position of the minima and maxima as well as the relative
magnitude of all features is preserved and only the amplitude is sensitive to radiation power.

Another confirmation of such behavior is illustrated in Fig.~\ref{Fig2}(b). This panel presents the experimental data at quite high intensity of $I\simeq 600$~W/cm$^2$ (thick red line) together with a fit (thin black line) using $\Delta\sigma_\text{osc}/\sigma= - A \exp(-\epsilon/f\tau_q)\epsilon\sin(2\pi\epsilon)$. The damping parameter $\tau_q=1.5$~ps is taken from similar fitting of the low-intensity data, obtained for $I\simeq 0.11$~W/cm$^2$ on the same sample, Ref.~\onlinecite{herrmann:2016}. Despite almost four orders of magnitude larger intensity, the shape of MIRO including the low-$B$ damping remains precisely the same, such that the only intensity-dependent parameter is the $B$-independent prefactor $A(I)$.

\begin{figure}
\includegraphics[width=0.9\linewidth]{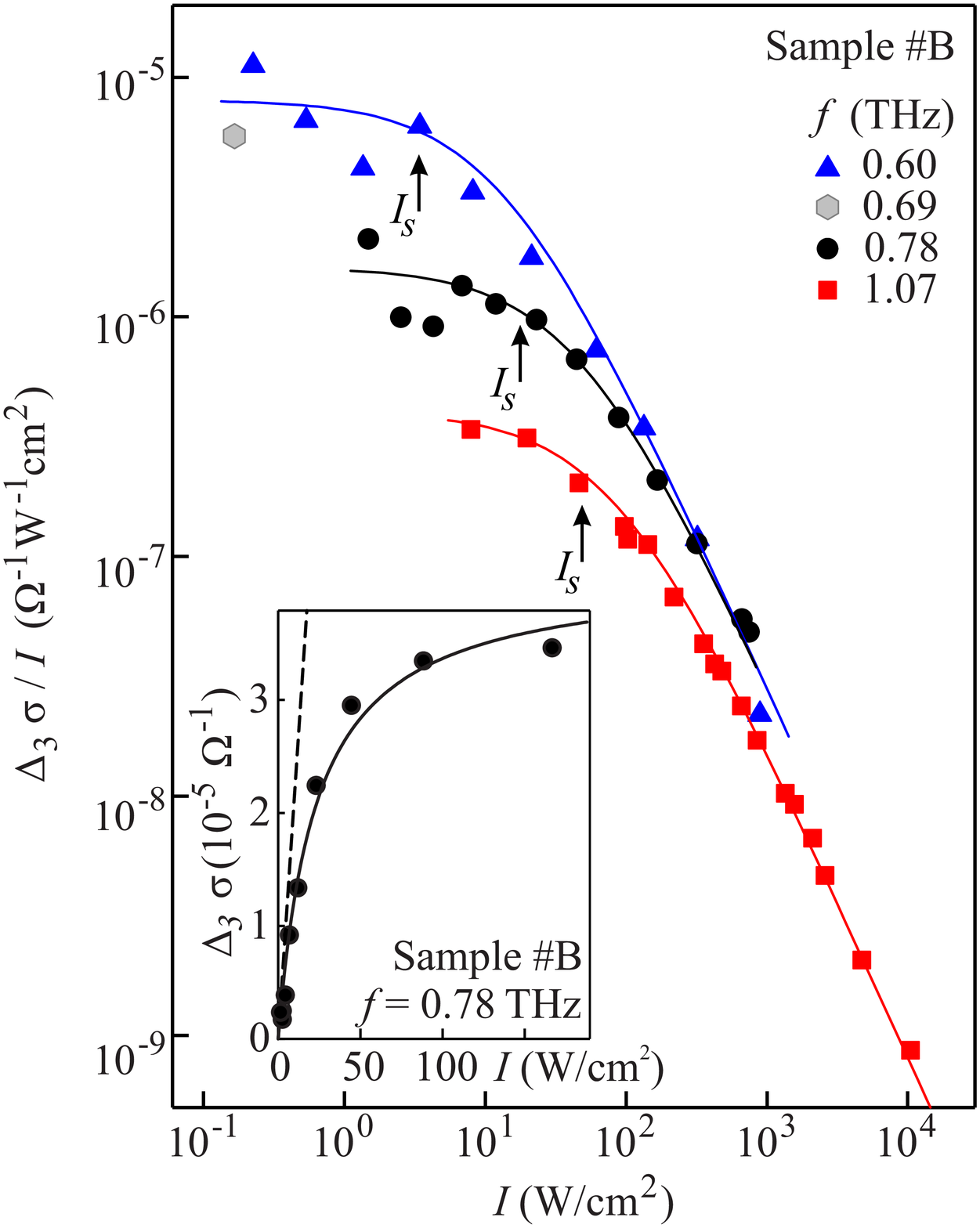}
\caption{Intensity dependence of the reduced oscillation amplitude $\Delta_3\sigma/I$ at the oscillation order $\epsilon =3$
for sample \#B, see definition in the text. The hexagon presents the result of Ref.~\onlinecite{herrmann:2016}
 obtained using low power $cw$ laser operating at $f=0.69$~THz.
Other data points correspond to excitation by THz laser pulses at three different frequencies.
The inset shows intensity dependence of the non-normalized oscillation amplitude $\Delta_3\sigma$
in a linear-linear plot. Solid lines are fits using $\Delta_3\sigma/I \propto (1 + I / I_s )^{-1.3}$.
Saturation intensities used for the fits are indicated by arrows and are given by $I_s=15$~W/cm$^2$ ($f = 0.6$~THz),
$I_s=45$~W/cm$^2$ ($f = 0.78$~THz), and $I_s=85$~W/cm$^2$ ($f = 1.07$~THz). The dashed line in the inset corresponds to a linear fit $\Delta_3\sigma\propto I$.
}
\label{Fig5}
\end{figure}

\begin{figure}
\includegraphics[width=0.9\linewidth]{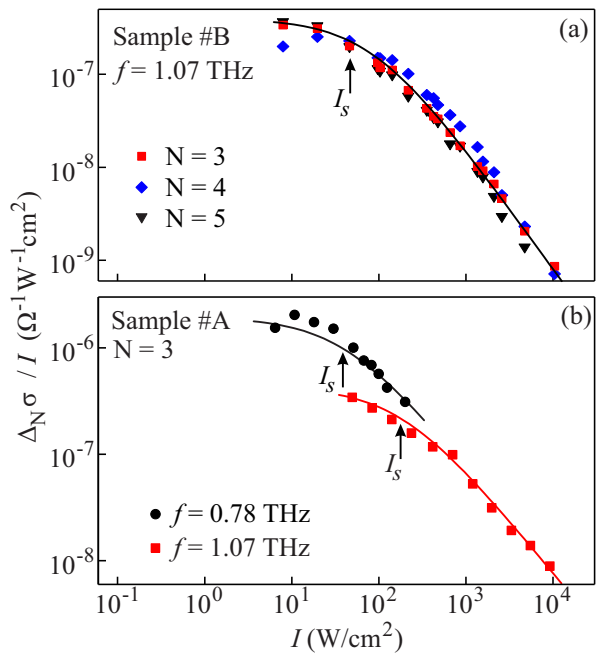}
\caption{Intensity dependence of the reduced oscillation amplitude $\Delta_N\sigma/I$. Solid lines are fits using $\Delta_3\sigma/I \propto (1 + I / I_s )^{-\beta}$. Panel (a) presents $\Delta_N\sigma/I$ for different oscillation orders, $N=$3, 4, and 5, measured on sample \#B under $f = 1.07$ THz illumination. Solid line is the fit using $\beta=1.3$ and $I_s=85$~W/cm$^2$. Panel (b) presents $\Delta_3\sigma/I$ measured on sample \#A under $f = 0.78$ and 1.07~THz radiation. The solid lines are fits using $\beta=1$ and saturation intensities $I_s= 30~(190)$~W/cm$^2$ for $f = 0.78~(1.07)$~THz.
}
\label{Fig6}
\end{figure}

To quantify the intensity dependence of the amplitude of MIRO, we extract 
the difference of
the maximal and minimal values of $\Delta \sigma$ near certain node at $\epsilon=N=3,\,4$ or 5,
and divide the resulting amplitude $\Delta_N\sigma$ by the intensity. The reduced amplitude, 
$\Delta_N\sigma/I$, as a function of intensity for samples \#B and \#A is
shown in Figs.~\ref{Fig5} and \ref{Fig6} for different oscillation orders $N$ and frequencies.
Double-logarithmic plots enable analysis for $I$ varying by five orders of magnitude while
the normalization by intensity enables easier comparison of the MIRO magnitude at different frequencies.
In the inset of Fig.~\ref{Fig5} the amplitude $\Delta_3\sigma$, not divided by intensity,
is presented in a linear-linear plot. This plot shows that the MIRO amplitude grows linearly with raising
power at low intensities and starts to saturate at higher intensities. For comparison, a hexagon in Fig.\,\ref{Fig5} presents the result of Ref.~\onlinecite{herrmann:2016} obtained using low power $cw$ laser operating at $f=0.69$~THz.

Figure~\ref{Fig5} shows the reduced amplitude $\Delta_3\sigma/I$ 
for sample \#B at three
radiation frequencies. In all three cases the amplitude can be well fitted by an empirical expression
\begin{equation}\label{intensity}
\Delta_N\sigma/ I = (1+I/I_s)^{-\beta},
\end{equation}
see solid lines in Fig.\,\ref{Fig5}.
The parameter $\beta$ is found to be $\beta\sim 1.3$ for sample $\#B$ and $\beta\sim 1$ for sample $\#A$ (see Fig.\,\ref{Fig6}), while the saturation intensity $I_s$ in all cases increases substantially with radiation frequency. Furthermore, while at low power (linear regime) MIRO  exhibit a strong frequency dependence, comparison of the traces for different frequencies at high intensities shows that the MIRO amplitude becomes almost frequency-independent in the saturation regime of the photoresponse. The observed intensity dependence is universal: it holds for different samples and oscillation orders, see Figs.~\ref{Fig6}(a) and (b), respectively.
In particular, Fig.~\ref{Fig6}(a) demonstrates that the reduced amplitude of MIRO, near different $\epsilon=3,\,4,\,5$ at fixed frequency,
can be well fitted by using one and the same saturation intensity $I_s$. This is another confirmation that the low-$B$ damping of oscillations in our experiments is insensitive to the intensity of radiation. One also observes that, for any given frequency, the saturation intensity for sample \#A is several times larger than for sample \#B, see Figs.~\ref{Fig5} and~\ref{Fig6}(b).

\begin{figure}
\includegraphics[width=0.9\linewidth]{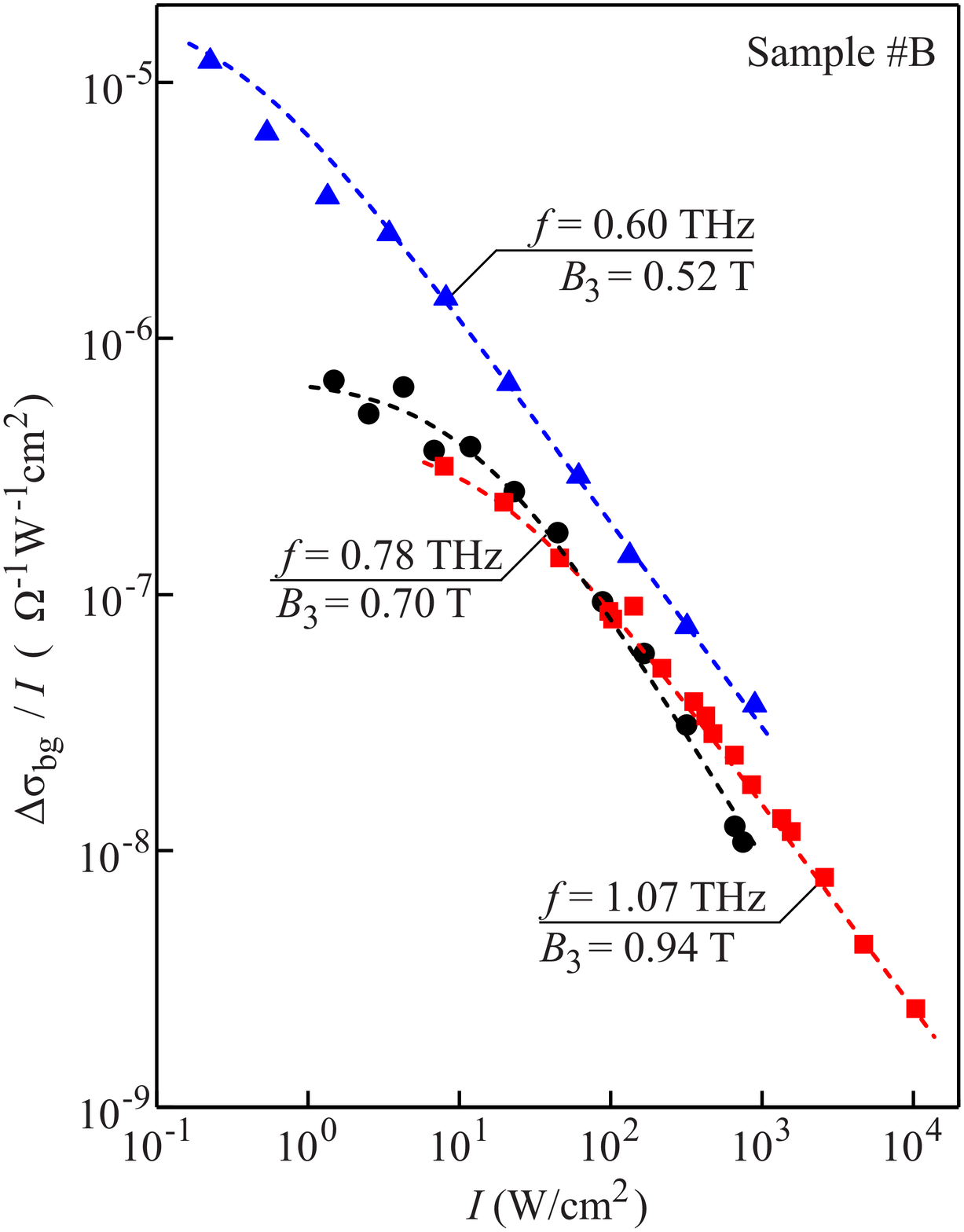}
\caption{The background part $\Delta\sigma_\text{bg}$ of the photoconductivity normalized to the radiation intensity $I$
measured for sample \#B at three radiation frequencies. Dashed lines are guide for the eye.
}
\label{Fig7}
\end{figure}

Sublinear intensity dependence is also detected for the background $\Delta\sigma_\text{bg}/\sigma$, see Fig.\,\ref{Fig7}.
As discussed above, the background can be extracted as a smooth line connecting $\Delta\sigma$ at positions of the MIRO nodes at integer $\epsilon$.
Figure~\ref{Fig7} presents the dependence of
$\Delta\sigma_\text{bg} / I$ measured
in the vicinity of $\epsilon = 3$, corresponding to $B=B_3$ in the figure.
It is seen that, compared to MIRO, the background becomes nonlinear at much lower intensities
(linear regime could even not be reached with our signal-to-noise ratios) and its behavior can not be described by Eq.~(\ref{intensity}). At large $I$, the background almost saturates. More precisely, at highest intensities the data in Fig.~\ref{Fig7} correspond to $\Delta\sigma_\text{bg} \propto I^\gamma$ with $0\leq\gamma\leq 0.2$.

\section{Discussion}
\label{discussion}

Summarizing our findings, we explored the evolution of MIRO over five decades of radiation intensity for several samples and several irradiation frequencies in the THz range.
Within the experimental accuracy, we find that MIRO as a function of inverse magnetic field $\epsilon=2\pi f/\omega_c\propto 1/B$ and intensity $I$ are under all conditions well described by the phenomenological equation,
\begin{equation}\label{overall}
\dfrac{\Delta\sigma_\text{osc}}{\sigma}=-A(I)\,e^{-\alpha\epsilon} \epsilon\sin(2\pi\epsilon),\quad A(I)\simeq\dfrac{cI}{(1+I/I_s)^\beta},
\end{equation}
with $\beta\sim 1$. This result is quite remarkable as the main features of oscillations such as shape and phase, as well as low-$B$ damping turn out to be roughly $I$-independent. The only quantity which depends on intensity is the overall prefactor $A(I)$, see Figs.~\ref{Fig5} and \ref{Fig6}.
As discussed below, the overall behavior described by Eq.~(\ref{overall}) is qualitatively different from nonlinear MIRO effects observed earlier in the microwave range of frequencies. 
Apart from the intensity dependence, we explored the sensitivity of high-power MIRO to edge/contact effects, as well as to the sense of circular polarization. This study establishes that high-intensity MIRO are governed by a bulk transport mechanism. Indeed, it was found previously \cite{herrmann:2016} that in the linear regime $\Delta\sigma\propto I$, illumination of the sample edges leads to a decrease of the MIRO amplitude roughly following the reduction of the illuminated area of the sample. The maximal signal was detected for illuminated spot not touching the edges. These measurements ruled out the contact/edge phenomena as a primary source of MIRO. However, these effects were not ruled out completely as they could simply be masked by stronger bulk effects.

Here we extended such studies to high intensity radiation. One can expect that the relative contribution of the edge/contact effects with respect to bulk effects may become larger at high power in comparison with low intensities.
Indeed, this necessarily happens if both effects are present, but the leading bulk one saturates earlier than the edge effect.
As seen in Fig.~\ref{Fig3}, no edge/contact contributions have been distinctly detected even for intensities strongly exceeding $I_s$. These results mean that if present, any edge/contact effects are  negligibly small as compared to the bulk effects. 
Therefore, in the following we discuss the results solely in terms of bulk mechanisms of MIRO.

Concerning the polarization dependence,
our previous study \cite{herrmann:2016} of MIRO induced by continuous low-power THz radiation demonstrated that, in the linear regime $\Delta\sigma\propto I$, oscillations are insensitive to the radiation helicity. These results, extending similar earlier findings \cite{smet:2005} in the microwave range, contradict the expectations based on the semiclassical Drude theory. 
Here we demonstrate that high-intensity MIRO (as well as the saturated background $\Delta\sigma_\text{bg}$) also feature immunity to the radiation helicity, see Figs.~\ref{Fig1}(a), \ref{Fig2}(a), and \ref{Fig3}. We note, however, that in the saturated regime $I\gg I_s$, the immunity is no longer surprising. One would expect it also in the case of helicity-dependent photoresponse. However, in this case the values of $I_s$ as well as MIRO at $I<I_s$ would be different for cyclotron-resonance-active and passive helicities, in contrast to our observations.

We turn now to a detailed discussion of the intensity dependence shown in Figs.~\ref{Fig5} and \ref{Fig6} and given by Eq.~(\ref{overall}). First of all, we notice that, despite the wide range of intensities studied, the $B$-positions of the minima and maxima of MIRO remain always the same and no additional structures appear in the high-power response, see Fig.~\ref{Fig4}. In other words, {\it the shape and phase of MIRO are found to be insensitive to $I$}, following $\Delta\sigma_\text{osc}\propto -\sin(2\pi\epsilon)$. As mentioned above, this behavior is qualitatively different from the nonlinear effects in the microwave range, where higher microwave power leads to emergence of additional minima and maxima \cite{zudov:2004,zudov:2006,dorozhkin:2007,pechenezhskii:2007,wiedmann:2009,lei:2006,dmitriev:2007,shi:2017}, or, at moderate power, to a shift of minima and maxima to the neighboring nodes at integer $\epsilon$ \cite{hatke:2011,vavilov:2004,dmitriev:2004}. The drastic differences between the high-power THz and microwave responses suggest distinct mechanisms of the nonlinearity.

As we argue below, in the terahertz range the major source of nonlinearity in the $\Delta\sigma_\text{osc}(I)$ dependence is {\it heating of electrons} in an illuminated sample. In this scenario, the elevated electron temperature $T_e(I)$ modifies the value of the amplitude but leaves the shape and phase unaffected, since $\sin(2\pi\epsilon)$ contains no quantities which are sensitive to $T_e$. Our estimates below (using both microscopic models of MIRO and results of previous work with low-intensity THz illumination\cite{herrmann:2016}) fully support this scenario and show that the nonlinearities found in the microwave range would require much higher intensities in the THz range, and may even become out of reach in the presence of strong heating, 
see Appendix~\ref{intrinsic} for details. One of the main reasons for that is that absorption of radiation responsible for heating is proportional to inverse square of the radiation frequency $f^{-2}$, while the MIRO amplitude scales as $f^{-4}$, see e.g. Ref.~\onlinecite{dmitriev:2012}.

Within the heating mechanism of nonlinearity, formulated above, not only the amplitude $A(I)$ but also the {\it low-$B$ damping} parameter $\alpha$ may acquire an implicit intensity dependence through $T_e(I)$. Such nonlinearity in $\alpha$ was identified before in the microwave range and was attributed there to the thermal broadening of Landau levels \cite{hatke:2011}.
Here we find, however, that  the parameter $\alpha$ in the empirical expression (\ref{overall}) does not depend on intensity, see also Fig.~\ref{Fig4}, and agrees well with the values $\alpha=1/f\tau_q\sim 1$ extracted for low intensity in Ref.~\onlinecite{herrmann:2016}.

In order to show that the thermal broadening of Landau levels is indeed negligible in our experiments,
we estimate the rate of inelastic electron-electron collisions, $\tau_\text{in}\simeq \hbar E_F/T_e^2$, where $E_F$ is the Fermi energy. This gives $\tau_\text{in}\sim 0.2$~ns for sample \#A and  $\tau_\text{in}\sim 0.15$~ns for sample \#B at $T_e=T_0\equiv4.2$~K. The associated high-power correction to $\alpha$ can be estimated \cite{dmitriev:2005,dmitriev:2009} as $\alpha(I)-\alpha(0)=1/f\tau_\text{in}(T_e)-1/f\tau_\text{in}(T_0)=a(T_e^2/T_0^2-1)$, where $a=1/f \tau_\text{in}(T_e=T_0)$. Since the resulting $a\lesssim 10^{-2} $ is pretty small [$a=4.7\; (6.2)\times 10^{-3}$ for sample \#A (\#B) and $f=1.07$~THz], it is not surprising that we detect no changes in $\alpha$ with intensity at moderate $I$.
Note that, according to the above estimate, the heating effect on $\alpha$ should become noticeable at higher intensities, provided the temperature reaches values $T_e\gtrsim 10~T_0$. This suggests that such temperature is not reached despite about two decades of intensity spanned in the nonlinear (saturation) regime of our experiments. Such situation is quite possible since heating is usually strongly nonlinear. In order to be compatible with the observed $\alpha\simeq\text{const}(I)\sim 1$ and the estimate $a\sim 10^{-2}$ above, the electron temperature should grow as $T_e\propto\sqrt{I}$ or slower in the nonlinear regime of heating.

Analysis of the data above shows that the intensity dependence of the {\it amplitude of MIRO} is well described by $A(I)\simeq c I(1+I/I_s)^{-\beta}$ with $\beta\sim 1$. In our previous experiments \cite{herrmann:2016}, corresponding to $I\ll I_s$ and thus to $A(I)=c I$, a good agreement was found between the measured values of the amplitude $A$
and its estimates obtained within inelastic mechanism of MIRO \cite{dmitriev:2003,dmitriev:2005}. In particular, experimental data for sample \#A illuminated by $f=0.69$~THz radiation of intensity $I=0.11$~W/cm$^2$ at $T=4.2$~K gave $A=0.056$ very close to the theoretical estimate $A=0.04$
obtained without fitting parameters \cite{herrmann:2016}.
Assuming that the inelastic mechanism also governs the nonlinear response, we can figure out whether the observed intensity dependence is compatible with the temperature dependence of parameters entering the theory. This theory predicts $A\propto I \tau_\text{in}(T_e) \tau_p^{-2}$. Here $\tau_p^{-1}$ is the momentum relaxation rate and $\tau_\text{in}^{-1}$, addressed above, describes the thermalization rate which balances changes in the energy distribution of electrons induced by radiation of intensity $I$. 
Taking  $\tau_p$ to be $T_e$-independent, we find that saturation of $A(I)\propto I/T_e^2$ at $I\gg I_s$ requires $T_e\propto I^{\beta/2}\sim \sqrt{I}$ in this regime. While in general, the sublinear dependendence of electron temperature on radiation intensity is well established for low-dimensional semiconductor systems at low temperatures~\cite{Ganichevbook,ch5Gantmakher1984,Ihn,ch5Karpus86p6,ch5Xie86p283,ch5Beregulin90p1138},
the feasibility of the particular dependence $T_e\propto\sqrt{I}$ requires additional study which is out of scope of the present work.

Note that in principle one cannot exclude that the momentum relaxation rate $\tau_p^{-1}$ also becomes $I$-dependent at high intensities, for instance, due to nonequilibrium phonon effects.
Inclusion of such effects into the theory is not straightforward, in particular, if typical phonon energies become comparable to $\omega_c$.
On a qualitative level it is reasonable to expect, however, that the non-elasticity of scattering processes may only reduce the amplitude of nonequilibrium oscillations in the distribution function responsible for MIRO. Furthermore, $\tau_p$ may only decrease with the increase of $I$ which would lead to an increase of $A$. Assuming $\tau_p\propto I^{-\eta}$ with $\eta>0$ at highest $I$ one obtains from $A(I)\sim\text{const}(I)$ that $T_e\propto I^{1/2-\eta}$, i.e. a stronger slow-down of heating at high intensities.

\section{Conclusion}

In conclusion, we have explored the magneto-resistance oscillations induced by high-power terahertz radiation with intensities ranging from 0.1~W/cm$^2$ to $\sim 10^4$~W/cm$^2$. For all studied AlGaAs/GaAs samples, all frequencies, and all oscillation orders we find a similar non-linear dependence of the amplitude of oscillations on intensity $I$, which follows an empirical function $I/(I+I/I_s)^\beta$ with $\beta\sim 1$. The sample-dependent saturation intensity $I_s$ is of the order of $10$~W/cm$^2$ and increases by $\sim 6$ times as the frequency $f$ raises from 0.6 to 1.1 THz. Additional experiments with the focused laser beam scanned over the sample reveal that no signal comes from the sample edges, even at highest intensities corresponding to strongly saturated bulk photoresponse. This shows that the observed oscillations originate in all studied regimes  from electrons in the bulk of the 2DES. 

We attribute the observed nonlinearity to heating effects within the inelastic mechanism of MIRO which was previously shown to dominate over the displacement mechanism in our samples \cite{herrmann:2016}. We observe that this scenario fits well our observations and suggests that the elevated electron temperature scales as $T_e\propto \sqrt{I}$ in the saturated regime of heating. A direct experimental access to the electron temperature would allow a more quantitative and critical test for this explanation. In sharp contrast to the microwave range of frequencies, where elevated power of illumination usually leads to emergence of a complex structures with multiple new minima and maxima, we detect no change in the shape of oscillations despite several decades of intensity change in the nonlinear regime. We argue that the main reason for such a drastically distinct behavior is the different frequency scaling of absorption (which is responsible for heating and scales as $f^{-2}$) and MIRO amplitude  (which scales as $f^{-4}$ and is further reduced due to heating effects). 
\vspace*{5mm}

\acknowledgments 
\label{acknow}

The support from the DFG (projects GA501/14-1 and DM1/4-1), 
the Volkswagen Stiftung Program and 
Russian Foundation for Basic Research (Grant No. 17-02-00384)
is gratefully acknowledged.

\appendix

\section{MIRO for illuminated and non-illuminated samples}
\label{appendixillum}
\begin{figure}
\includegraphics[width=0.9\linewidth]{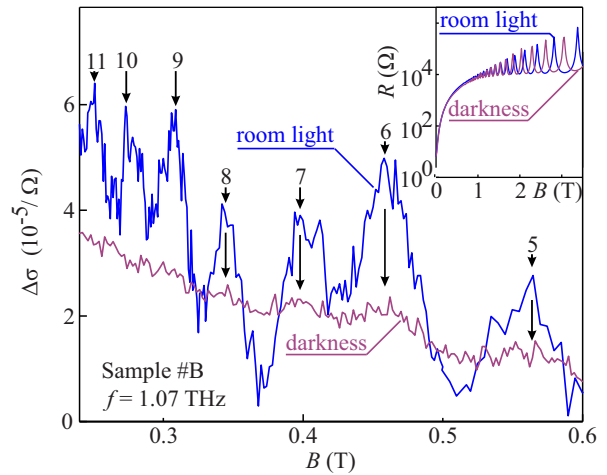}
\caption{Terahertz radiation-induced photoconductivity
as a function of magnetic field as measured in sample \#B in
the dark at $I = 300$~W/cm$^{2}$ and after illumination with room light at $I = 40$~W/cm$^{2}$.
The data for the dark conditions are obtained at higher intensities to enable resolving of MIRO.
The inset shows corresponding magneto-transport data. }
\label{fig8}
\end{figure}

Figure~\ref{fig8} shows terahertz radiation induced oscillations measured for the sample \#B
in the dark and after illumination with room light.  Comparison of these traces demonstrates
that while weak MIRO can be identified in the former case
they are much better pronounced in the latter one. A more detailed analysis of the role of illumination is relegated to future studies.

\section{MIRO for different geometries of Corbino samples}
\label{appendixgeometry}

\begin{figure}
\includegraphics[width=0.9\linewidth]{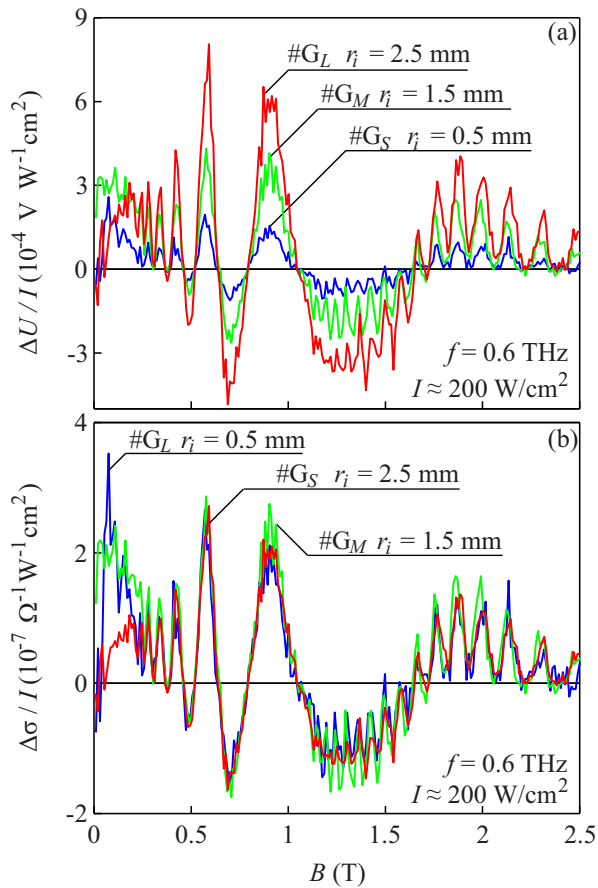}
\caption{Magnetic field dependence of the photo-induced voltage drop signal $\Delta U$ (panel a) and the photoconductivity $\Delta\sigma$ (panel b),
both normalised to radiation intensity $I$, for three samples \#G$_L$, \#G$_M$, \#G$_S$. The samples are made from the same wafer but have different inner Corbino radius, $r_i$, equal to 2.5, 1.5, and 0.5~mm, respectively.
}
\label{Fig9}
\end{figure}

To explore a possible effect of the sample geometry on MIRO signal we prepared three
samples from the same wafer, with the same diameter of the outer contact but with different diameters
of inner contacts.
The results are shown in Fig.~\,\ref{Fig9}.
The figure demonstrates that the detected voltage signals are higher for the samples with larger diameter of the
inner contacts. The photoconductivity, however, taking into account the geometrical factor $1/(2\pi) \ln (r_o/r_i)$ shows
almost the same deoendence for $r_i = 0.5$ mm and $r_i = 1.5$ mm. 
A somewhat smaller magnitude of $\Delta\sigma$ detected for sample \#G$_L$ having the largest $r_i = 2.5$~mm
is attributed to the decrease of the radiation power because for this geometry
a substantial part of the sample is covered by non-transparent gold film of the central electrode.

\section{Insignificance of intrinsic nonlinear effects}
\label{intrinsic}

We now discuss our results in light of the {\it nonlinear MIRO} effects observed previously in the microwave range of frequencies. Phenomenologically, these effects can be classified as follows, according to the type of changes they produce in the shape and amplitude of MIRO.

{\it Fractional MIRO} is characterized by additional maxima and minima emerging around certain fractional harmonics of the cyclotron resonance, most prominently, around $\epsilon=\omega/\omega_c=1/2,\,3/2,$ and $2/3$, where the minima could even develop into fractional zero resistance states at higher power. These effects were extensively studied both experimentally \cite{zudov:2004,zudov:2006,dorozhkin:2007,pechenezhskii:2007,wiedmann:2009} and theoretically \cite{pechenezhskii:2007,lei:2006,dmitriev:2007} and generally require stronger magnetic field, $\omega_c>\tau_q^{-1}$, such that the width of disorder-broadened Landau levels is comparable to or smaller than the distance between them.

In conditions $\omega_c\lesssim \tau_q^{-1}$ of overlapping Landau levels relevant to our experiment, a transition to a sublinear power dependence of the amplitude of MIRO accompanied by a progressive shift of the positions of minima and maxima to the closest nodes at integer $\epsilon$ was observed \cite{hatke:2011} at a moderate microwave power. Very recent experiments \cite{shi:2017} in extremely high quality sample using high-power low-frequency microwave source revealed pronounced {\it fine structure of MIRO} with multiple minima and maxima around integer $\epsilon=N$ which emerge at high power and shift to $\epsilon=N$ as power is further increased. These phenomena were successfully explained within the mainstream theoretical framework combining the displacement\cite{ryzhii:1970,ryzhii:1986,durst:2003,vavilov:2004,khodas:2008} and inelastic\cite{dmitriev:2003,dmitriev:2004,dmitriev:2005} mechanisms of MIRO and were attributed to a combination of multiphoton and nonlinear feedback effects in the corresponding parameter regimes \cite{shi:2017,hatke:2011,vavilov:2004,dmitriev:2004,khodas:2008}.

On top of the above {\it intrinsic} mechanism-specific effects which strongly affected the shape of magnetooscillations, experiments in the microwave range revealed traces of radiation-induced heating of electrons \cite{hatke:2011}. Within the same theoretical framework, the electron temperature does not enter MIRO explicitly. The heating may only modify the parameters, in particular those sensitive to temperature-dependent inelastic processes.

Taking into account the above, the behavior summarized in Eq.~(\ref{overall}) suggests the major role of heating effects and, at the same time, relative weakness of the intrinsic nonlinear effects which would otherwise change the shape of oscillations. This situation becomes possible if a characteristic intensity $I_i$ marking the onset of intrinsic effects (in the absence of heating) substantially exceeds the characteristic intensity $I_s$ for the onset of heating effects.

To establish whether the condition $I_i\gg I_s$ can indeed be met in our experiments, we first recall that absorption of radiation (responsible for heating) is proportional to inverse square of the radiation frequency, leading to $I_s\propto f^{2}$. In turn, the MIRO amplitude in the regime $\Delta\sigma\propto I$ is predicted to scale as $f^{-4}$, which gives $I_i\propto f^4$. This fact alone makes the situation $I_i\gg I_s$  at THz frequencies quite feasible, given that $I_i \approx  I_s$ in Ref.~\onlinecite{shi:2017} and taking into account that the radiation frequencies used there were about $50$ times smaller than in our work.

A more precise estimate of the ratio $I_i/I_s$ can be obtained from the analysis of MIRO under low-power continuous THz illumination performed in our previous work \cite{herrmann:2016}. Using the results obtained there, for $f=1.07$~THz and $T=4.2$~K we estimate $I_i^{\#B} \approx  0.1~I_i^{\#A}
\approx  1$~kW/cm$^2$. This intensity is indeed much larger than $I_s \approx  40$~W/cm$^2$ observed in the experiment, which confirms that intrinsic nonlinearities are not relevant at $I \approx  I_s$. Since the characteristic intensity for intrinsic nonlinearities is itself sensitive to electron temperature, for $I_i\gg I_s$ the intensity $\tilde{I_i}$, at which intrinsic nonlinearities start to play a role, should be found self-consistently. For example, assuming ${\tilde I}_i\propto T_e^2$ as in the inelastic mechanism of MIRO with $T_e$-independent $\tau_p$, one should expect $\tilde{I}_i\approx  10^2~I_i$ for $T_e(\tilde{I}_i) \approx 10~T_0$. Since in our experiments both relations $I_i\gg I_s$ and $\tilde{I}_i\gg I_i$ are expected to be met, it is not surprising that intrinsic nonlinearities were not detected despite high intensities up to $10$~kW/cm$^2$.

\end{document}